\documentclass{aa}
\usepackage{amsfonts}
\usepackage{amssymb}
\usepackage{graphicx}
\usepackage{natbib}
\begin{document}
   \title{Massive zero-metal stars: Energy production and mixing}

   \author{C. W. Straka
          \and
          W. M. Tscharnuter
          }

   \offprints{C. W. Straka,\\ \email{cstraka@ita.uni-heidelberg.de}}

   \institute{Institut f\"ur Theoretische Astrophysik,
     Universit\"at Heidelberg, Tiergartenstra{\ss}e 15,
     69121 Heidelberg, Germany\\
   }

   \date{Received; accepted}

   \abstract{Time-dependent nuclear network calculations at
     constant temperature show that for zero-metal stars
     $\gtrsim 20 M_\mathrm{\sun}$ (i) $\beta$-decay reactions and
     (ii) the  $^{13}$N(p,$\gamma$)$^{14}$O reaction
     must be included. It is also shown that the nuclear timescale in these
     zero-metal stars is shorter than the  mixing timescale and therefore the
     assumption of instantaneous mixing across convective regions is not
     fulfilled. We conclude that proper modeling of these processes may
     alter the evolution of massive zero-metal stars.
   }

   \maketitle
%

\section{Introduction}
The evolution of zero-metal stars in the mass range
$\sim 20$--$100 M_\mathrm{\sun}$
, i.e., massive Population III stars (Pop-III for short),
has been studied since the
pioneering work of \citet{ezer71} who followed the evolution of stars
in the mass range $5$--$100 M_\mathrm{\sun}$ from the
pre-main sequence contraction phase until the exhaustion of
hydrogen on the main
sequence. This opened the field for many studies that were mainly concerned
with later evolutionary stages
\citep{cary74, castellani83, eleid83, ober83, klapp83, klapp84}.
Motivated by the still ongoing debate about the initial mass function
of Pop-III stars,
\citet{marigo01} present the most comprehensive study of zero-metal
evolutionary
models ($0.7$--$100 M_\mathrm{\sun}$) starting from the ZAMS
until the AGB in the case of low- and intermediate-mass stars,
or to the onset of carbon burning in massive stars.  

In earlier studies the authors were forced to make numerous assumptions
about equilibria between the chemical species involved.
The most recent studies by \citet{marigo01}, however,
incorporate nuclear networks without making any assumptions about equilibria.
Nevertheless, even these recent state-of-the-art studies rely on some
simplifications that are certainly fulfilled during the evolution of
\emph{normal} stars but are questionable in the case of massive zero-metal
stars.

It is the aim of this letter to show that the following two assumptions widely
used, namely
\begin{enumerate}
\item $\beta$-decay negligible against proton-capture
\item instantaneous mixing across convective regions
\end{enumerate}
are not met during the evolution of massive Pop-III stars and that
proper modeling may alter the evolution of these objects.


\section{Nuclear Network}
\begin{figure*}
  \resizebox{\textwidth}{!}
  {\includegraphics[angle=-90]{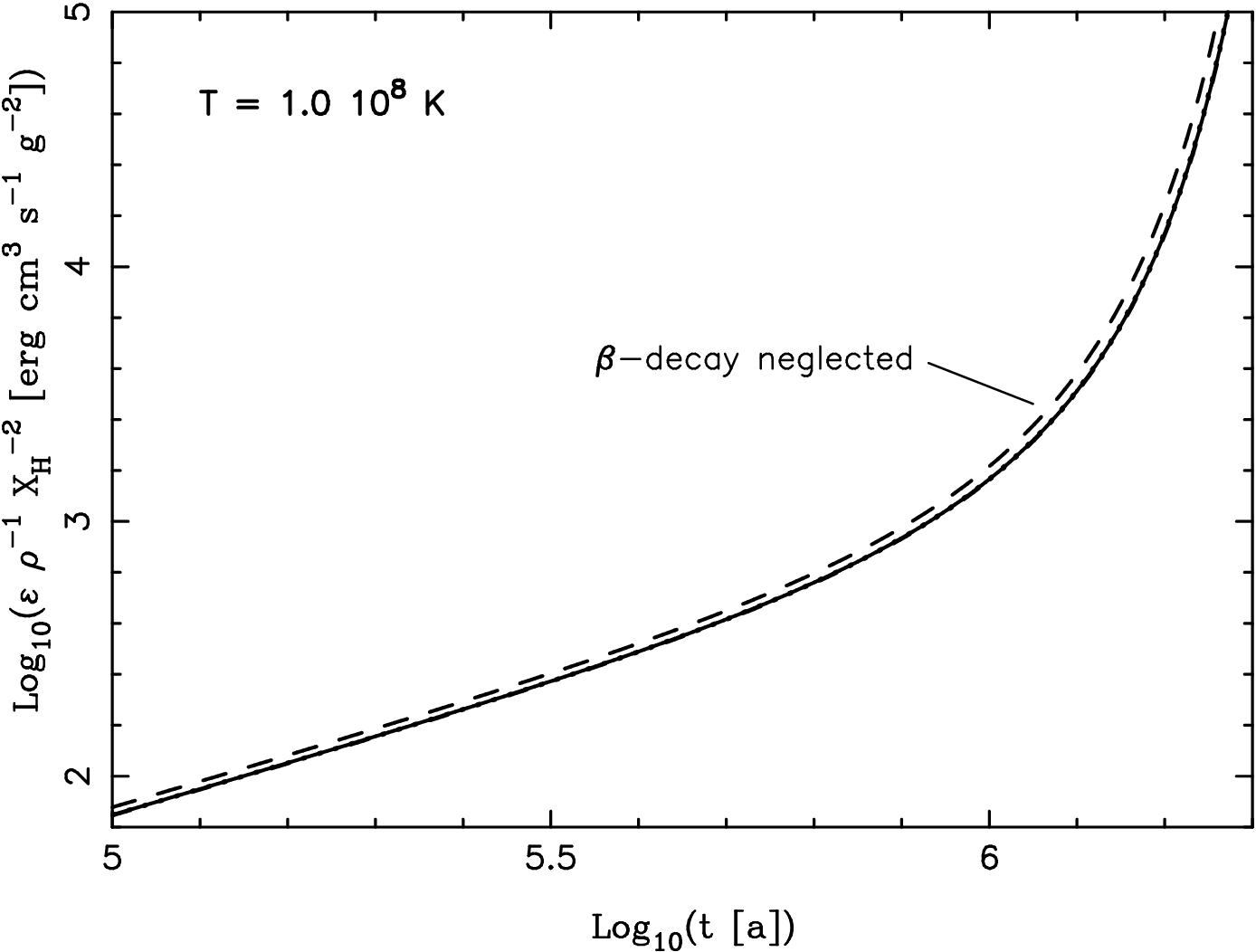}
  \hspace{1cm}
  \includegraphics[angle=-90]{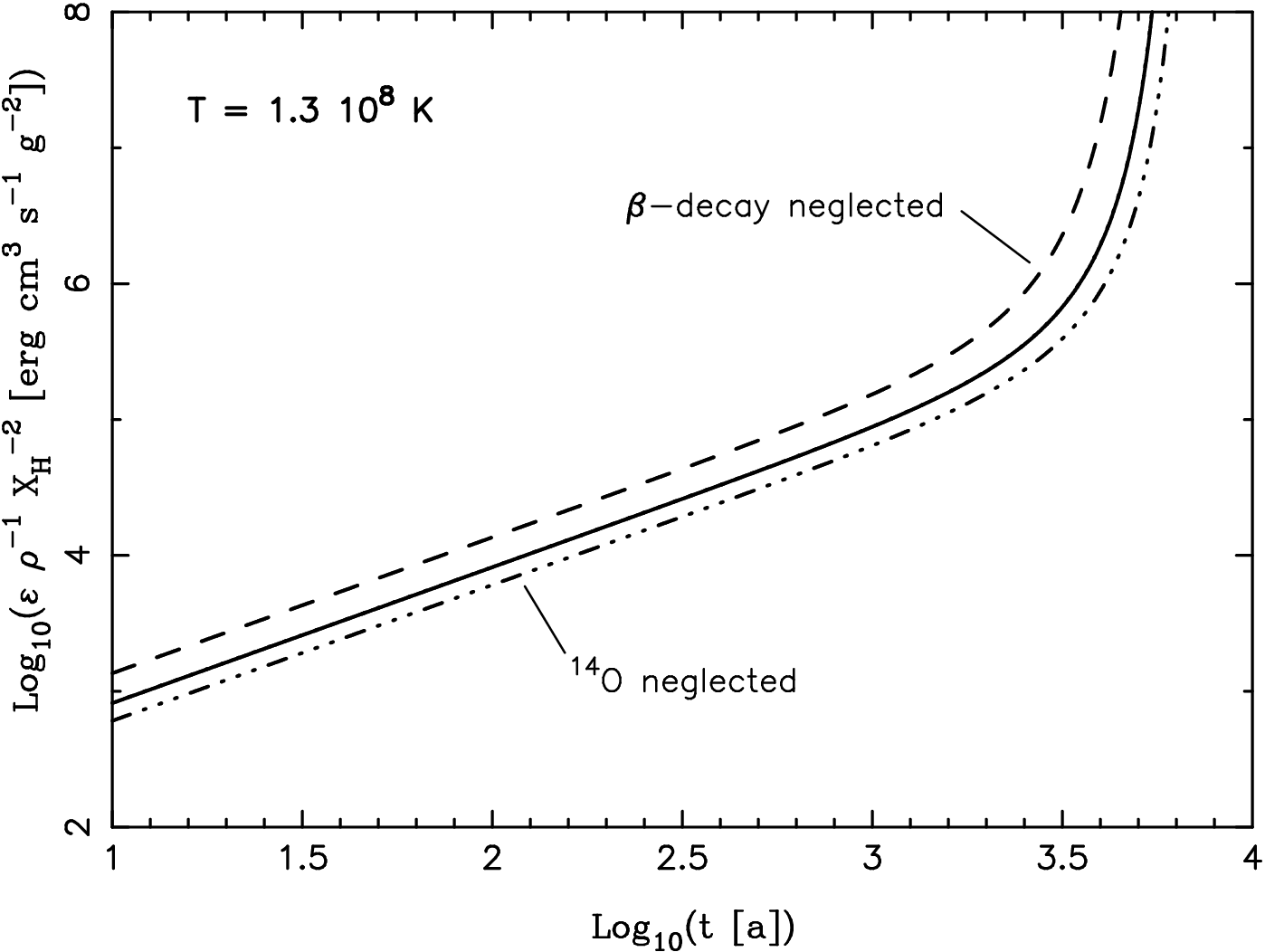}}
  \caption{Energy generation  per $\rho X_\mathrm{H}^2$ at constant
    temperature for {\bf a}) $T = 1.0\, 10^8 \mathrm{K}$
    and {\bf b}) $T = 1.3\, 10^8 \mathrm{K}$. The solid curve is our
    reference calculation including all 15 chemical species.
    The dashed curve neglects the $\beta$-decay and the dotted-dashed curve
    neglects the reaction $^{13}$N(p,$\gamma$)$^{14}$O. Differences between
    the reference curve and the curve neglecting
    $^{13}$N(p,$\gamma$)$^{14}$O are too small to show up in {\bf a}).}
  \label{fig:eprod}
\end{figure*}
It was first noted by \citet{ezer61} that Pop-III stars more massive than
about $20 M_\mathrm{\sun}$ are supplied by a different mode of energy
generation. Due to the lack of the elements C, N and O the CNO-cycle cannot
generate the luminosity needed to halt contraction. This raises the
temperature to the point where the $3\alpha$-reaction produces
enough carbon to initiate the CNO-cycle. As a result, the CNO-cycle
operates at a much higher temperature, typically at $10^8\, \mathrm{K}$.
Since the energy generation in the CNO-cycle exhibits a strong
temperature dependence the central temperatures of the more massive
models up to $100 M_\mathrm{\sun}$ do not exceed 
$T_\mathrm{c} = 1.3 \, 10^8\, \mathrm{K}$
\citep[see e.g.,][]{eleid83, marigo01}. At these temperatures, the CNO-cycle
turn-over rate is governed by the beta-decay half-lives \citep{ezer71}.

This finding has one important consequence for modeling stellar evolution:
one has to include the elements $^{13}$N and $^{15}$O (mono-cycle),
$^{17}$F (bi-cycle) and $^{18}$F (tri-cycle) explicitely in time-dependent
network calculations. In addition,
for temperatures exceeding $10^8\, \mathrm{K}$ the proton
capture $^{13}$N(p,$\gamma$)$^{14}$O 
(followed by $^{14}$O($\beta^{+}$,$\nu$)$^{14}$N) supersedes the
 $^{13}$N($\beta^{+}$,$\nu$)$^{13}$C rate. Contrary to the findings of
\citet{klapp83} our calculations show a non-negligible effect on the energy
generation for $T > 10^8\, \mathrm{K}$.

In order to estimate the described effects, we have performed time-dependent
network calculations at constant temperature. Our reference calculation
includes 15 chemical species: $^{1}$H, $^{4}$He, $^{12}$C,
$^{13}$C, $^{13}$N, $^{14}$N, $^{15}$N, $^{14}$O, $^{15}$O, $^{16}$O,
$^{17}$O, $^{18}$O, $^{17}$F, $^{18}$F and $^{19}$F with initial abundances of
$X_\mathrm{H} = 0.77$, $X_\mathrm{He} = 0.23$ and
$X_\mathrm{other} = 10^{-15}$, and the following reaction chains:
\begin{itemize}
\item $3\alpha$-process
\item CNO tri-cycle
\item $^{13}$N(p,$\gamma$)$^{14}$O and $^{14}$O($\beta^{+}$,$\nu$)$^{14}$N
\quad\mbox{.}
\end{itemize}
Reaction rates are taken from
\emph{NACRE}\footnote{http://pntpm.ulb.ac.be/Nacre/} \citep[see][]{angulo99}. 
The calculations are performed utilizing the DAE solver \emph{LIMEX} that
is maintained by \citet{ehrig98}.

Our results can be viewed in Fig.~\ref{fig:eprod} where we have plotted the
energy generation per $\rho X_\mathrm{H}^2$ against the time in years for the
two cases $T = 1.0 \, 10^8\, \mathrm{K}$ (see Fig.~\ref{fig:eprod}a) and
$T = 1.3 \, 10^8\, \mathrm{K}$ (see Fig.~\ref{fig:eprod}b). Neglecting
the beta-decay half-lives (dashed curves) has an impact on the energy
generation of $\sim 10\%$ for $1.0 \, 10^8\, \mathrm{K}$ and
$\sim 65\%$ for $1.3 \, 10^8\, \mathrm{K}$. Whereas the proton capture on
$^{13}$N alters the energy generation at $1.0 \, 10^8\, \mathrm{K}$
only slightly ($< 1\%$), it has a $\sim 30\%$ effect at
$1.3 \, 10^8\, \mathrm{K}$ (dotted-dashed curve).
It is worth noting that the energy generation is only marginally ($< 1\%$)
effected by both the second (named $^{17}$ON) and the third cycle
(named $^{19}$FO). These cycles need only be included
if one is interested in the relative abundances of the elements O and F.
At constant temperature, the energy production \emph{never} attains
a constant value due to the ongoing feeding of freshly produced carbon via
the $3\alpha$-process. 


\section{Mixing}
Massive stars contain large convective cores in which 
mixing of chemical species occurs due to the turbulent convective motion.
In \emph{normal} stars, this process is very rapid compared to the
slow changes of the chemical composition produced by nuclear reactions.
Under these circumstances one can safely assume that the composition
in a convective region always remains homogeneous, i.e., elements are
instantaneously mixed over the convective region. 
The following arguments are put forward to show that the turn-over time
of the CNO-cycle at the high temperatures of massive Pop-III stars
can be comparable to the timescale of turbulent mixing -- or even shorter.

\subsection{Mixing timescale} 
We show that the mixing timescale, $\tau_\mathrm{conv}$, is of the
order of 10 days in the case of both normal and zero-metal stars. An estimate
of this timescale is easily derived from the standard mixing length
theory \citep[MLT,][]{vitense58} and the formulas we present resemble those
given in \citet[chap.~7]{kippenhahn90}. Let us assume a standard composition
of $X_\mathrm{H} = 0.77$ and $X_\mathrm{He} = 0.23$, i.e., $\mu = 0.58$,
and opacity due to electron scattering $\kappa = 0.2 (1 + X_\mathrm{H})$,
i.e., $\kappa = 0.35\,\mathrm{cm^2\, g^{-1}}$.
Even in the case of hot Pop-III stars, neglecting
radiation pressure does not change this order of magnitude estimate
and it suffices to assume a monoatomic ideal gas:
$\delta = 1$, $c_\mathrm{P} = 5 \mathfrak{R} / 2 \mu$.
The mixing timescale is approximately given by the pressure
scale height $H_\mathrm{P}$ divided by the velocity of
convective motion $v_\mathrm{conv}$. With this:
\begin{equation}
\tau_\mathrm{conv} \sim 10^{8} \left(\frac{r}{R_\mathrm{\sun}}\right)^2
\left(\frac{M_\mathrm{\sun}}{m}\right)
\sqrt{\frac{T}{10^8\, \mathrm{K}}}
\left(\frac{10^{-4}}{\sqrt{\nabla - \nabla_\mathrm{e}}}\right)\, \mathrm{s}
\label{eq:tauconv}
\end{equation}
where $r$, $T$ and $m$ are the radius, temperature and mass,
respectively, at one particular
locus in a star. The quantity $\nabla - \nabla_\mathrm{e}$ is the difference
between the actual temperature gradient, $\nabla$, and $\nabla_\mathrm{e}$
which describes the variation of $T$ in a mass element during its motion.
Introducing two dimensionless variables:
\begin{eqnarray}
U &\sim& 2\, 10^{-11} \left(\frac{m}{M_\mathrm{\sun}}\right)
\left(\frac{T}{10^8\, \mathrm{K}}\right)^{3/2} \times \nonumber \\
&& \times \left(\frac{100\, \mathrm{g\,cm^{-3}}}{\rho}\right)^2
\left(\frac{R_\mathrm{\sun}}{r}\right)^2\,\mbox{,}
\label{eq:u} \\
W &=& \nabla_\mathrm{rad} - \nabla_\mathrm{ad}\,\mbox{,}
\label{eq:w}
\end{eqnarray}
and provided that $U \ll (\nabla - \nabla_\mathrm{e})^{1/2} \ll W$ 
(see below) the cubic equation of MLT can be simplified yielding
\begin{equation}
\sqrt{\nabla - \nabla_\mathrm{e}} \sim \left(\frac{8}{9} U W \right)^{1/3}
\,\mbox{.}
\label{eq:nne}
\end{equation}
For an estimate of $\tau_\mathrm{conv}$ at the \emph{center} of a star
the first two equations can be even further reduced, since
$m = 4/3\mathrm{\pi} \rho_\mathrm{c} r^3$:
\begin{eqnarray}
\tau_\mathrm{conv}  &\sim& 10^{6}
\left(\frac{M_\mathrm{\sun}}{m}\right)^{1/3}
\left(\frac{100\, \mathrm{g\, cm^{-3}}}{\rho_\mathrm{c}}\right)^{2/3} 
\left(\frac{T_\mathrm{c}}{10^8\, \mathrm{K}}\right)^{1/2} \nonumber \\
&& \times \left(\frac{7\, 10^{-4}}{\sqrt{\nabla - \nabla_\mathrm{e}}}\right)
\,\mbox{s}
\label{eq:tausimple}
\end{eqnarray}
and
\begin{equation}
U \sim 4\, 10^{-10} \left(\frac{m}{M_\mathrm{\sun}}\right)^{1/3}
\left(\frac{100\, \mathrm{g\, cm^{-3}}}{\rho_\mathrm{c}}\right)^{4/3}
\left(\frac{T_\mathrm{c}}{10^8\, \mathrm{K}}\right)^{3/2}
\label{eq:usimple}
\end{equation}

For a typical example, consider a $20 M_\mathrm{\sun}$ Pop-III star. There
we have $T_\mathrm{c} \sim 10^8 \mathrm{K}$,
$\rho_\mathrm{c} \sim 100\,\, \mathrm{g\, cm^{-3}}$.
Let us choose a mass coordinate: $m = 1 M_\mathrm{\sun}$, hence
$U \sim 4\, 10^{-10}$ from Eq. (\ref{eq:usimple}). For reasonable values
of $W$, $W \sim 1 \ldots 100$ it follows that
$(\nabla - \nabla_\mathrm{e})^{1/2} \sim 7\, 10^{-4}$. Note
that the assumption  $U \ll (\nabla - \nabla_\mathrm{e})^{1/2} \ll W$
is fulfilled. With Eq. (\ref{eq:tausimple}) we finally arrive at
\begin{equation}
\tau_\mathrm{conv} \sim 10\, \mathrm{days}\,\mbox{.}
\end{equation}
For a star containing metals the central temperature is by a factor of three
smaller but as normal stars are less compact also the density goes
down by a factor of ten. Thus $\tau_\mathrm{conv}$ changes by a factor
of two or so. Similar results are obtained at other loci of the convective
core and for stellar masses in the range $20$--$100\, M_\mathrm{\sun}$.

\subsection{Nuclear timescale}
\begin{figure}
  \centering
  \resizebox{\hsize}{!}
  {\includegraphics[angle=-90]{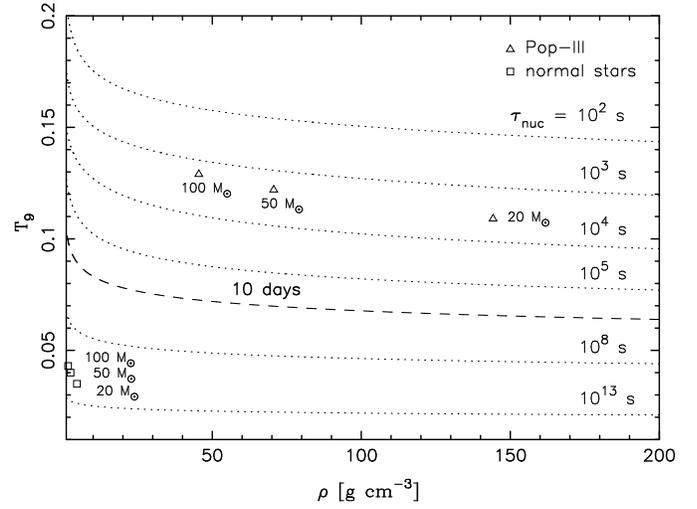}}
  \caption{Lines of constant $\tau_\mathrm{nuc}$ at $X_\mathrm{H} = 0.77$.
    In the region \emph{above} the dashed line the mixing time is \emph{slow}
    against the nuclear timescale.}
  \label{fig:taunuc}
\end{figure}
A characteristic timescale for nuclear reactions in massive stars is the
turn-over time of the CNO-cycle. This time is primarily given by the
slowest reaction of the cycle which is $^{14}$N(p,$\gamma$)$^{15}$O.
Even in the high temperature regime ($T = 1.3\, 10^8 \mathrm{K}$)
where proton-captures become comparable to $\beta$-decays
this estimate holds within a factor of two. Hence,
\begin{equation}
\tau_\mathrm{nuc} \sim \tau_\mathrm{n14pg} = \frac{A_\mathrm{H}}{X_\mathrm{H}
\rho \lambda_\mathrm{n14pg}(T)}\,\mbox{,}
\end{equation}
where $A_\mathrm{H}$, $X_\mathrm{H}$ and $\rho$ are the atomic mass of
hydrogen, the mass abundance of hydrogen and the density,
respectively \citep[see][]{clayton83}. The reaction
rate $\lambda_\mathrm{n14pg}(T)$ measured in
[$\mathrm{cm^3\, Mol^{-1}\, s^{-1}}$] is only a function of temperature.
Taking this reaction rate again from \emph{NACRE} \citep[see][]{angulo99}:
\begin{eqnarray}
\lambda_\mathrm{n14pg}(T) &=& 
         4.83\, 10^7\, T_\mathrm{9}^{-2/3} \times \nonumber \\
         && \times \exp\left(-15.231\, T_\mathrm{9}^{-1/3} 
           - \left(\frac{T_\mathrm{9}}{0.8}\right)^2\right) \nonumber \\
         && \times (1 - 2.00\, T_\mathrm{9} 
         + 3.41\, T_\mathrm{9}^2 - 2.43\, T_\mathrm{9}^3) \nonumber \\
         &+& 2.36\, 10^3\, T_\mathrm{9}^{-3/2}\,
         \exp(-3.010\, T_\mathrm{9}^{-1}) \nonumber \\
         &+& 6.72\, 10^3\, T_\mathrm{9}^{0.380}\,
         \exp(-9.530\, T_\mathrm{9}^{-1})\,\mbox{,}
\end{eqnarray}
the nuclear timescale is easily calculated for a given set of density $\rho$
and temperature $T$. To explore the parameter space we plot
lines of constant time $\tau_\mathrm{nuc}$ in a $\rho-T$ diagram
(see Fig. \ref{fig:taunuc}). The dashed line marks the typical timescale
of convective mixing. Thus, the region \emph{above (below)} this line is
characterized by \emph{slow (fast)} mixing compared to the nuclear timescale.
The assumption of instantaneous mixing is only justified in the lower
region where mixing is fast. It can be nicely seen that all \emph{normal}
stars lie in the lower region but that massive Pop-III stars lie in
the upper region where the assumption of instantaneous mixing does not hold.

%
\section{Conclusions}
Performing time-dependent nuclear network calculations at
constant temperature we show that in the case of massive Pop-III stars
($\gtrsim 20\, M_\mathrm{\sun}$) neglecting the $\beta$-decay against
proton-capture leads to a considerable error in the energy generation rate.
In addition, the reaction $^{13}$N(p,$\gamma$)$^{14}$O cannot be omitted for
temperatures exceeding $10^8\, \mathrm{K}$.

Moreover, the nuclear timescale of massive Pop-III stars
can be very short (order of hours) compared to the timescale of
convective mixing which is of the order of 10 days. Therefore,
instantaneous mixing which is well justified in normal stars
may introduce large errors in evolutionary calculations of zero-metal
stars.  

Contrary to normal stars massive Pop-III stars have not forgotten their
nu\-cle\-ar his\-to\-ry since equi\-lib\-ri\-um abun\-dances be\-tween
chemical species are never attained. Hence, evolutionary calculation
must start on the pre-main sequence well before the onset of
nuclear reactions.

\begin{acknowledgements}
  We thank W. J. Duschl for helpful discussions and for
  improving the manuscript. We are particularly grateful to
  R. Ehrig for providing and
  supporting LIMEX. Part of this work was supported by the German
      \emph{Deut\-sche For\-schungs\-ge\-mein\-schaft, DFG\/}
      (SFB 439 Galaxies in the young universe).
\end{acknowledgements}

\end{document}